\documentclass[apjl,iop,twocolappendix,numberedappendix]{emulateapj}

\usepackage{graphicx}
\usepackage{amsmath}

\newcommand{\sigsfr}{\Sigma_\mathrm{SFR}}
\newcommand{\siggas}{\Sigma_\mathrm{gas}}
\newcommand{\tff}{t_\mathrm{ff}}
\newcommand{\gastff}{(\siggas/t)_\mathrm{single-ff}}
\newcommand{\gastmff}{(\siggas/t)_\mathrm{multi-ff}}

\slugcomment{Accepted for publication in ApJ Letters, \today}

\shorttitle{A universal, turbulence-regulated star formation law}
\shortauthors{Salim, Federrath, Kewley}

\begin{document}

\title{A universal, turbulence-regulated  star formation law: from Milky Way clouds to high-redshift disk and starburst galaxies}

\author{Diane~M.~Salim\altaffilmark{1}, Christoph~Federrath\altaffilmark{1}, Lisa~J.~Kewley\altaffilmark{1}}
\email{christoph.federrath@anu.edu.au}

\altaffiltext{1}{Research School of Astronomy and Astrophysics, The Australian National University, Canberra, ACT~2611, Australia}

\begin{abstract}
Whilst the star formation rate (SFR) of molecular clouds and galaxies is key in understanding galaxy evolution, the physical processes which determine the SFR remain unclear. This uncertainty about the underlying physics has resulted in various different star formation laws, all having substantial intrinsic scatter. Extending upon previous works that define the column density of star formation ($\sigsfr$) by the gas column density ($\siggas$), we develop a new universal star formation (SF) law based on the multi-freefall prescription of gas. This new SF law relies predominantly on the probability density function (PDF) and on the sonic Mach number of the turbulence in the star-forming clouds. By doing so we derive a relation where the star formation rate (SFR) correlates with the molecular gas mass per multi-freefall time, whereas previous models had used the average, single-freefall time. We define a new quantity called \emph{maximum (multi-freefall) gas consumption rate} (MGCR) and show that the actual SFR is only about 0.4\% of this maximum possible SFR, confirming the observed low efficiency of star formation. We show that placing observations in this new framework ($\sigsfr$ vs.~MGCR) yields a significantly improved correlation with 3--4 times reduced scatter compared to previous SF laws and a goodness-of-fit parameter $R^2=0.97$. By inverting our new relationship, we provide sonic Mach number predictions for kpc-scale observations of Local Group galaxies as well as unresolved observations of local and high-redshift disk and starburst galaxies that do not have independent, reliable estimates for the turbulent cloud Mach number.
\end{abstract}

\keywords{galaxies: high-redshift --- galaxies: ISM --- galaxies: starburst --- ISM: clouds --- stars: formation --- turbulence}

\section{Introduction}
The evolution of galaxies is driven by the intricate power play between interstellar turbulence and gravity controlling the formation of stars in giant molecular clouds \citep{Ferriere:2001aa,MacLowKlessen2004,ElmegreenScalo2004,McKeeOstriker2007,PadoanEtAl2014}. The rate at which stars form is hence a pivotal quantity in tracing a galaxy's fundamental properties and distribution of matter and activity. Nonetheless, the functional dependence of the column density of star formation ($\sigsfr$) has been a highly debated topic for more than two decades, with historical parameterizations including the mean column density of available gas ($\siggas$) \citep{Schmidt1959,K98,BigielEtAl2008,LeroyEtAl2008,Daddi:2010aa,SchrubaEtAl2011,RenaudEtAl2012,KennicuttEvans2012}, as well as the ratio between $\siggas$ and the mean freefall time $\tff$ \citep{KDM12,Federrath:2013aa,Krumholz2014}. However, significant scatter remains in both these approaches, such that $\sigsfr$ can vary by more than an order of magnitude for any given input $\siggas$ or $\siggas/\tff$. Follow-up analyses of these theoretical models via computer simulations have determined that the observed scatter may be primarily attributed to the physical variations in the sonic Mach number ($\mathcal{M}$) of the turbulence in the star-forming clouds \citep{Federrath:2013aa}. This results in separate $\mathcal{M}$-dependent relations for the SFR, which cover the observed range of $\sigsfr$. The aim of this work is hence to unify these separate laws and develop a unique, universal relation.

In Section~\ref{sec:theory} we introduce and derive our new SF law based on the PDF and Mach number of interstellar turbulence. Section~\ref{sec:obs} describes the observational sample utilized to assess our new SF law. In Section~\ref{sec:results} we present our findings and compare our new model to previous parameterizations. In Section~\ref{sec:appl} we invert our new SF law to make sonic Mach number predictions for extragalactic sources. Finally, Section~\ref{sec:conclusions} summarizes our conclusions.

\section{A new star formation rate descriptor} \label{sec:theory}

Here we derive an improved SF law, which takes into account the distribution of gas densities (PDF) present in galactic clouds and the interstellar medium (ISM). This follows from the finding that the observed scatter in the SF law can be principally attributed to the variation in the sonic Mach number of the clouds and galaxies observed \citep{Federrath:2013aa}.

\subsection{The probability density function (PDF) of interstellar gas}
In 2012 Krumholz, Dekel and McKee initiated the refinement of the traditional Kennicutt-Schmidt law by dividing the gas surface density by the freefall time of the gas ($\siggas/\tff$), indeed achieving a stark improvement in star formation rate correlation \citep{KDM12,Federrath:2013aa,Krumholz2014}. However, because this model implements only the division between the \emph{average} surface density and the \emph{average} freefall time of the gas, important information regarding the wide distribution of densities within the ISM and molecular clouds \citep{KainulainenEtAl2009,SchneiderEtAl2012,SchneiderEtAl2013,KainulainenFederrathHenning2013,Kainulainen:2014aa} is not taken into account. We hence extend upon the important previous work by \citet{KDM12} to incorporate the density PDF, with the ultimate goal of eliminating the remaining scatter in the previous relation.

In our new SF law, we must consider the fact that denser gas forms stars at a higher rate \citep{KrumholzMcKee2005,PadoanNordlund2011,Hopkins2013IMF} because gas with higher density $\rho$ has a shorter freefall time than more diffuse gas,
\begin{equation} \label{eq:tff}
\tff(\rho) = \sqrt{\frac{3\pi}{32G\rho}},
\end{equation}
where $G$ is the gravitational constant. Note that the concept of a multi-freefall time was pioneered by \citet{HennebelleChabrier2011,HennebelleChabrier2013} and \citet{ChabrierEtAl2014}. It suggests that, due to the clumpy nature of molecular clouds, the typical timescale for star formation is not the average timescale of the clouds, but the \emph{density-dependent} timescale of each collapsing substructure within the clouds. This \emph{multi-freefall concept} has recently received support from numerical simulations in \citet{FederrathKlessen2012} and from observations in \citet{EvansEtAl2014}.

In contrast, \citet{KDM12} and \citet{Federrath:2013aa} had only correlated $\sigsfr$ with $\siggas(\rho_0)/\tff(\rho_0)$ (hereby denoted as $\gastff$); i.e. only at a \emph{single} freefall time evaluated at the mean density $\rho_0$. It is conceivable, however, that the true correlation must be in the form of an integral over the whole density PDF, $\int[\siggas(\rho)/\tff(\rho)]\,p_\rho d\rho$. In the following, we compute this integral by making the standard assumption that the PDF follows a log-normal distribution as the initial condition for star formation,
\begin{equation} \label{eq:pdf}
p(s)ds=\frac{1}{\sqrt{2\pi\sigma_s^2}}\exp{\left(-\frac{(s-s_0)^2}{2\sigma_s^2}\right)}ds,
\end{equation}
where the density is written in logarithmic form, normalized to the mean density
\begin{equation} \label{eq:s}
s=\ln(\rho/\rho_0).
\end{equation}
This transformation of variables from $\rho$ to $s$ has advantages during the integration step below and allows us to write the PDF, Equation~(\ref{eq:pdf}), in its standard log-normal form. The mean logarithmic density $s_0$ is related to the variance $\sigma_s^2$ by $s_0=-\sigma_s^2/2$ \citep{Vazquez1994}. As derived by \citet{PadoanNordlund2011} and \citet{MolinaEtAl2012}, the logarithmic density variance is given by
\begin{equation} \label{eq:sigs}
\sigma_s^2 = \ln\left(1+b^2\mathcal{M}^2\frac{\beta}{\beta+1}\right),
\end{equation}
parameterized by the sonic Mach number $\mathcal{M}$, the turbulent driving parameter $b$ \citep{FederrathKlessenSchmidt2008,FederrathDuvalKlessenSchmidtMacLow2010}, and the ratio of thermal to magnetic pressure, plasma $\beta$.

The log-normal PDF, Equation~(\ref{eq:pdf}), and the associated density variance -- Mach number relation, Equation~(\ref{eq:sigs}), are observed to provide a good approximation to the real PDF in molecular clouds \citep{KainulainenEtAl2009,Brunt2010,SchneiderEtAl2012,SchneiderEtAl2013,Kainulainen:2014aa}, in the Galactic centre \citep{RathborneEtAl2014}, on large Galactic scales \citep{BerkhuijsenFletcher2008}, and even in extra-galactic systems \citep{HughesEtAl2013,BerkhuijsenFletcher2015}.

\subsection{Derivation of the maximum (multi-freefall) gas consumption rate (MGCR)} 

For a rigorous derivation, we start by defining a cartesian coordinate system in which the line-of-sight toward the cloud or galaxy is in the $z$-direction and maps of column density and star formation rate surface density are in the $xy$-plane. First, the gas column density is defined as $\siggas(\rho)=\int_0^{H_z}\rho dz$, where $H_z$ is the scale height of the cloud or galaxy for which $\siggas$ is to be determined. From observations and simulations discussed in the previous section, we know that $\siggas$ and $\rho$ follow closely a log-normal distribution, Equation~(\ref{eq:pdf}). Thus, in order to compute an average of a density-dependent variable, we simply have to integrate that variable over the entire PDF. For the average gas column density, for example, we would evaluate $\int\siggas p_{\Sigma\mathrm{gas}} d\siggas = \iint\rho p_\rho d\rho dz = \iint \rho_0 \exp(s) p(s) ds dz = \rho_0 \int dz = \rho_0 H_z$, which is indeed the average column density $\siggas(\rho_0)$. Now we follow exactly the same mathematical procedure, but for the combined density-dependent variable $\siggas/\tff=\siggas(\rho_0)/\tff(\rho_0)\exp(3s/2)$, i.e.,
\begin{align}
\gastmff & = \int_0^{\infty} \frac{\siggas(\rho)}{\tff(\rho)}\,p_\rho d\rho \\
                                                        & = \frac{\siggas(\rho_0)}{\tff(\rho_0)} \int_0^{\infty} (\rho/\rho_0)^{3/2}\,p_\rho d\rho \\
                                                        & = \frac{\siggas(\rho_0)}{\tff(\rho_0)} \int_{-\infty}^{\infty} \exp\left(\frac{3}{2}s\right)p(s)ds \\
                                                        & = \frac{\siggas(\rho_0)}{\tff(\rho_0)} \exp\left(\frac{3}{8}\sigma_s^2\right). \label{eq:sol}
\end{align}
Note that in the second step, we use the exact scaling $\siggas/\tff\sim\rho^{3/2}$, because $\siggas\sim\rho$ and $\tff\sim\rho^{-1/2}$. The third step transforms variables from $\rho$ to $s$ via Equation~(\ref{eq:s}) and via the identity $p_\rho d\rho = p(s)ds$. The last step, Equation~(\ref{eq:sol}), is the analytic solution of the integral over all densities.

We call this new quantity derived above, the \emph{maximum gas consumption rate} or \emph{multi-freefall gas consumption rate} (MGCR), denoted $\gastmff$.

\subsection{The single-freefall to multi-freefall correction factor}

We see that Equation~(\ref{eq:sol}) includes the multi-freefall correction factor given by $\exp(3\sigma_s^2/8)$, which corrects the single-freefall average $\siggas(\rho_0)/\tff(\rho_0)\equiv\gastff$ used in \citet{KDM12} and \citet{Federrath:2013aa} for the underlying PDF of densities.

We can now insert the density variance from Equation~(\ref{eq:sigs}) into Equation~(\ref{eq:sol}) and obtain the multi-freefall correction factor,
\begin{equation} \label{eq:correction}
C_\mathrm{multi-ff} \equiv \frac{\gastmff}{\gastff} = \left( 1+b^2\mathcal{M}^2\frac{\beta}{\beta+1} \right)^{3/8}.
\end{equation}

\begin{table*}
\caption{Summary of observational sources considered in deriving and testing our new SF law}
\def\arraystretch{1.6}
\setlength{\tabcolsep}{4.3pt}
\begin{tabular}{llcccccc}
\hline
Data Source                & Source Type/Name      & $\log_{10}\sigsfr$  & $\log_{10}\siggas$    & $\log_{10}(\frac{\Sigma_\mathrm{gas}}{t})$ & $\mathcal{M}$ & Correction  & $\log_{10}(\frac{\Sigma_\mathrm{gas}}{t})$ \\

& & & & $\hspace{2.0em}_\mathrm{single-ff}$ & Range & Factor & $\hspace{2.0em}_\mathrm{multi-ff}$ \\

\hline
\citet{BolattoEtAl2011}          & Small Magellanic Cloud (SMC) & $-2.15$&$1.10$&$-0.670$&  16--200&$12.6$&0.429\\
\hline
\citet{YusefZadeh2009}     & Central Molecular Zone (CMZ) &$-0.750$&$2.08$&$0.900$&  50 &$9.47$&1.88\\
                        
\hline
\citet{HeidermanEtAl2010}        & Taurus                    &$-0.750$&$2.04$&$1.60$& 10   &$1.11^\star$&1.64\\
(H10)              & C2D + GB Clouds           &$-0.135$&$1.89$&$1.68$& 5--20&$2.89$&2.14\\
                        & Class I YSOs              &$\phantom{-}0.208$&$2.24$&$2.40$& 1--2&$1.06$&2.43 \\
                        & Class I YSOs (upper limit)&$-0.147$&$2.01$&$2.15$& 1--2&$1.06$&2.17 \\
                        & Flat SED YSOs             &$\phantom{-}0.332$&$2.28$&$2.47$& 1--2&$1.06$&2.50\\
                        & Flat SED YSOs(upper limit)&$-0.090$&$2.00$&$2.10$& 1--2&$1.06$&2.13\\
\hline                       
\citet{WuEtAl2010} (W10)   & HCN(1-0) Clumps           &$\phantom{-}1.17$&$2.95$&$3.46$  & 2--5&$1.40$&3.60\\                        
\hline                      
\citet{GutermuthEtAl2011}        & Class II YSO counts in  & \\
(G11)              & eight molecular clouds    &$-0.787$&$1.64$&$1.04$& 5--20&$2.89$&1.50\\
\hline
\citet{LadaLombardiAlves2010}             & Molecular clouds at $A_k \geq 0.1$ &$-0.941$&$1.46$&$0.722$& 5--20&$2.89$&1.18 \\
(L10)              & Molecular clouds  at $A_k \geq 0.8$ &$\phantom{-}0.545$&$2.37$&$2.57 $& 5--20&$2.89$&3.03 \\
\hline
\end{tabular}
\begin{minipage}{\linewidth}
\vspace{0.1cm}
\textbf{Notes.} Column~1: Reference. Column~2: Source classification. Column~3: Geometric mean of $\Sigma_{\mathrm{SFR}}$. Column~4: Geometric mean of $\Sigma_{\mathrm{gas}}$. Column~5: Geometric mean of $\gastff$. Column~6: $\mathcal{M}$ ranges whose geometric mean of the upper and lower limit were used to calculate the average Mach number. Column~7: Correction factor based on the average Mach number as attained in Equation~(\ref{eq:correction}). Note that we assume plasma $\beta\to\infty$ (i.e., no magnetic field correction) for all except Taurus' correction factor (see $\star$), for which we use the available measurement of $\beta=0.02$ from \citet{HeyerBrunt2012}. Column~8: Our new SFR descriptor, $\gastmff$, called maximum (multi-freefall) gas consumption rate (MGCR), computed via Equation~(\ref{eq:correction}), and derived in Equation~(\ref{eq:sol}).
\end{minipage}
\label{tab:obs}
\end{table*}

\begin{figure}
\centerline{\includegraphics[width=\linewidth]{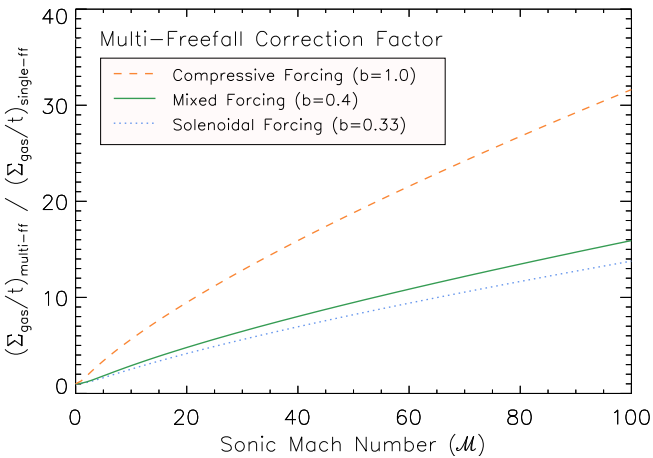}}
\caption{The multi-freefall correction factor derived in Equation~(\ref{eq:correction}) as a function of the sonic Mach number for three values of the turbulent driving parameter $b$ whilst neglecting magnetic fields ($\beta\to\infty$). The values in the figure represent the correction factor are applied to existing measurements of $\gastff$ in order to correct them to $\gastmff$.}
\label{fig:correction}
\end{figure}

Figure~\ref{fig:correction} shows the $\gastmff/\gastff$ correction factor, Equation~(\ref{eq:correction}), as a function of the Mach number $\mathcal{M}$ for three values of the turbulent driving parameter $b$ and---for simplicity---neglecting magnetic fields, i.e., $\beta\to\infty$. We see that the multi-freefall correction is always positive and increases with increasing Mach number. The correction factor is about an order of magnitude for typical cloud Mach numbers, \mbox{$\mathcal{M}\sim5$--$50$}, which means that our correction to the previous SF laws will be significant. In the following, we will assume a fixed turbulent driving parameter, $b=0.4$, representing a natural mixture. However, we emphasize that $b$ is not likely fixed across all molecular clouds \citep{Brunt2010,PriceFederrathBrunt2011,GinsburgFederrathDarling2013,Kainulainen:2014aa}, but in the absence of a direct measurement of $b$, we presume a standard value for all clouds.

\section{Observational Sample Selection} \label{sec:obs}

We selected our sample of 11 galactic objects and 1 extragalactic object based on the availability of gas or dust column density ($\siggas$) and SFR column density ($\sigsfr$).
Submillimeter observations are the primary source for the study of molecular gas clouds in galaxies. In combination with young stellar object (YSO) counts, infrared and ultraviolet luminosities, $\sigsfr$ measurements may be obtained. As our new SFR descriptor is primarily dependent on the sonic Mach number of the gas in each observed source, availability of such estimates were vital in selecting relevant data. Mach number estimates for Milky Way clouds \citep{HeidermanEtAl2010,LadaLombardiAlves2010,GutermuthEtAl2011}, molecular clumps \citep{WuEtAl2010}, YSO \citep{HeidermanEtAl2010} and the Central Molecular Zone (CMZ) \citep{YusefZadeh2009} as well as the Small Magellanic Cloud (SMC) \citep{BolattoEtAl2011} were taken from and are summarized in \citet{Federrath:2013aa}. A summary of the data and sonic Mach number estimates utilized are listed in Table~\ref{tab:obs}.

Derivation of $\gastmff$ was calculated via Equation~(\ref{eq:correction}) using $b=0.4$ and $\beta\rightarrow\infty$ for all cases except the Taurus molecular cloud. Taurus is known to exhibit high levels of magnetic activity \citep{HeyerBrunt2012}, so an estimate of $\beta=2\mathcal{M}_A^2/\mathcal{M}^2\sim0.02$ \citep[with the sonic and Alfv\'en Mach number, $\mathcal{M}=10$ and $\mathcal{M}_A=1$, respectively; see][]{HeyerBrunt2012} was applied to obtain Taurus' multi-freefall correction factor. 

Previous SF laws have reported data points attained from individual clouds and assigned a linear correlation between $\sigsfr$ and $\siggas$ or $\gastff$ \citep{KennicuttEvans2012,KDM12}. However, because we do not have sonic Mach number measurements for individual clouds, utilizing the data points from each cloud to derive a law from our new descriptor would not be an accurate method of derivation. We can only obtain approximate values of sonic Mach number for populations of clouds, so in order to arrive at a fair comparison for each relation we report the geometric mean values of $\sigsfr$, $\siggas$, $\gastff$, and $\mathcal{M}$ for each group of clouds and apply robust line fits to these data points. The error bars associated with each point were derived from calculating the standard deviation of the mean. Since the previous SFR descriptors $\siggas$ and $\gastff$ do not depend on the sonic Mach number of the clouds, uncertainty in the values of $\mathcal{M}$ was not propagated when deriving values of $\gastmff$ in order to objectively evaluate differences between our new SFR descriptor and previous works.

\section{Results: an improved star formation law} \label{sec:results}

\begin{figure*}
\centerline{\includegraphics[width=\linewidth]{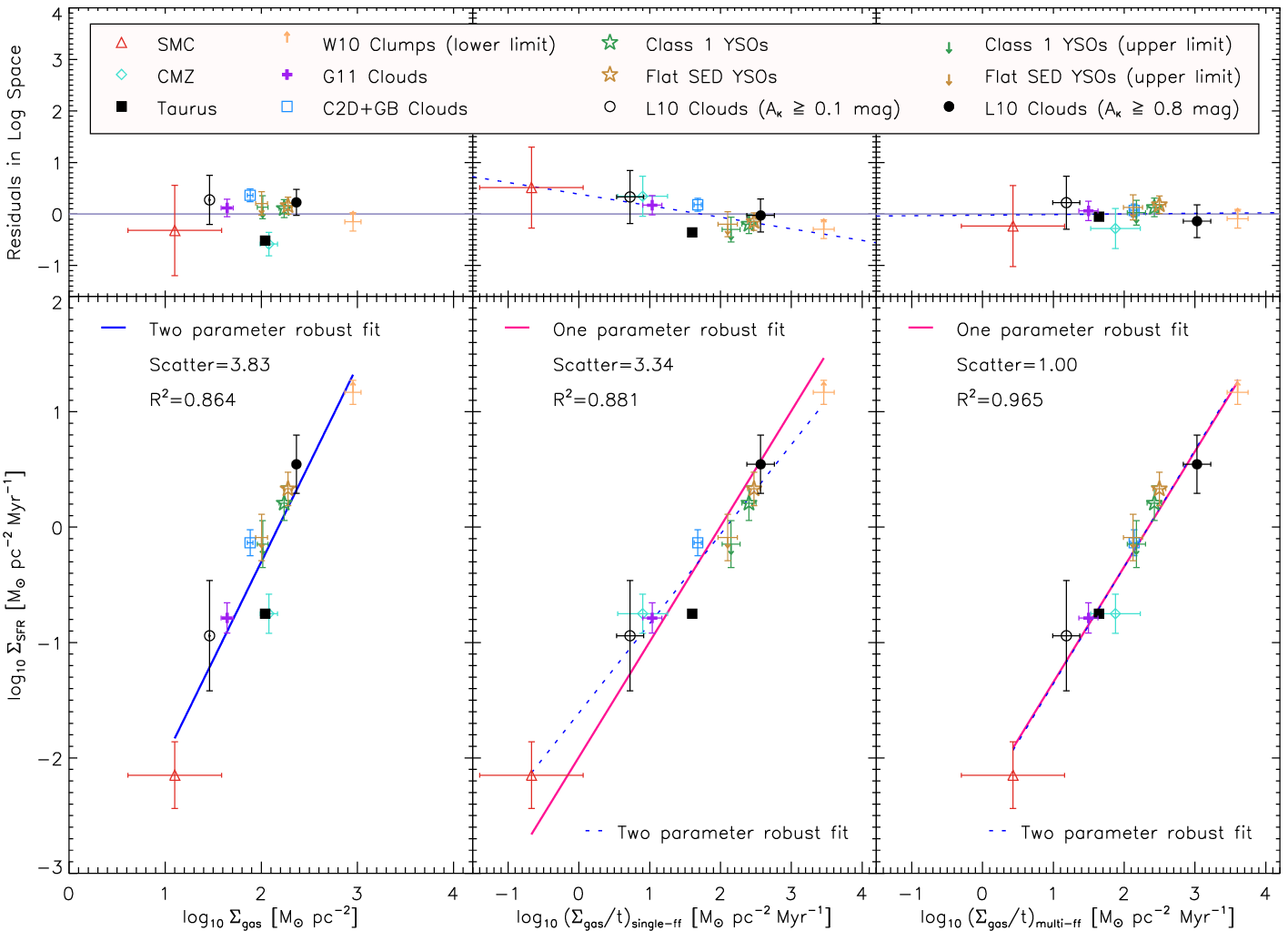}}
\caption{\emph{Bottom panels:} $\sigsfr$ versus $\siggas$ (classical Kennicutt-Schmidt relation; left), $\sigsfr$ versus $\gastff$ \citep[][middle]{KDM12}, and our new model, $\sigsfr$ versus $\gastmff$ (right) for the observational data in Table~\ref{tab:obs}. Power-law fits are shown as solid lines and the respective goodness-of-fit parameter $R^2$ (where $R^2=1$ corresponds to a perfect fit) is shown in each panel, as well as the normalized scatter. We also apply a two-parameter robust line fit to these relations, shown as the dotted lines. We find that our new multi-freefall SF law provides the best correlation with a reduced scatter by a factor of 3--4 compared to the previous SF laws. \emph{Top panels:} The residuals of each relation in log space versus each of the parameterizations being analyzed. The solid and dotted lines in each panel are respectively the fits shown in the bottom panels.}
\label{fig:sflaws}
\end{figure*}

Figure~\ref{fig:sflaws} shows a direct comparison of previous SF laws, $\sigsfr$ vs.~$\siggas$ (left panel) and $\sigsfr$ vs.~$\gastff$ (middle panel), together with our new MGCR correlator, $\sigsfr$ vs.~$\gastmff$ (right panel), including all the data listed in Table~\ref{tab:obs}. For the classical Kennicutt-Schmidt relation (left-hand panel), we apply a two-parameter robust power-law fit, with the offset and the slope of the power law as free parameters. For the \citet{KDM12} relation (middle panel) and for our new multi-freefall relation (right-hand panel), we respectively apply a one-parameter robust fit with the offset as the only free parameter (the slope is fixed to unity). Thus, both the \citet{KDM12} single-freefall SF law and our new multi-freefall SF law predict a direct linear relationship between the actual SFR and the maximum possible single-freefall or maximum possible multi-freefall gas consumption rate, respectively. We also apply the two-parameter robust fit to both these relations in order to gauge how much they deviate from the assumed unity slope.

\begin{table}
\caption{Summary of the results of the statistical tests applied to the data and the fits generated to model their relationship}
\def\arraystretch{1.6}
\setlength{\tabcolsep}{5.7pt}
\begin{tabular}{lcccc}
\hline
$\log_{10}\sigsfr$ correlator &  Offset & Slope & Scatter & $R^2$ \\
\hline
\multicolumn{5}{c}{Two-parameter fits} \\
$\log_{10}\siggas$   &  -3.70  & $1.70\pm0.21$       & 3.83     & 0.864 \\
$\log_{10}\gastff$   &  -1.61  & $0.78\pm0.05$       & 1.07     & 0.962 \\
$\log_{10}\gastmff$  &  -2.37  & $1.01\pm0.06$       & 0.99     & 0.965 \\  
\hline
\multicolumn{5}{c}{One-parameter fits} \\
$\log_{10}\gastff$   &  -1.99 & 1 (fixed)  & 3.34   & 0.881 \\
$\log_{10}\gastmff$  &  -2.34  & 1 (fixed) & 1.00   & 0.965 \\  
\hline
\end{tabular}
\begin{minipage}{\linewidth}
\vspace{0.1cm}
\textbf{Notes.} Column~1: SFR descriptor in log space. Column~2: Offset of the relation's robust line fit in log space. Column~3: Slope of the relation's robust line fit in log space. Column~4: Relative scatter of the data. Column~5: Goodness-of-fit, $R^2$ value ($R^2=1$ would indicate a perfect fit).
\end{minipage}
\label{tab:fits}
\end{table} 

In order to quantitatively evaluate the relative goodness of fit for each relation, we have applied two statistical tests to the data generated by each parameterization and their corresponding lines of best fits, summarized in Table~\ref{tab:fits}. The first value, which we define as the relative scatter in each relation, is the unweighted $\chi^2$ value normalized with respect to our new SFR descriptor, $\gastmff$. Scatter values greater than unity hence indicate a greater degree of scatter than that present in our new relation. The second test applied attains a value for the goodness-of-fit, $R^2$. The $R^2$ value is the coefficient of determination and has a range of 0 to 1. An $R^2=1$ indicates a model that perfectly fits the data. 

Both the statistical tests we have applied show a significantly tighter correlation between $\sigsfr$ and $\gastmff$ than with either $\gastff$ or $\siggas$, as summarized in Table~\ref{tab:fits}. For the $\gastff$ and $\gastmff$ parameterizations, this conclusion is further emphasized by the behavior of their residuals and gradients of their two parameter least-squares fits. The two-parameter fits suggest that there is significant intrinsic correlation remaining in the $\gastff$ parameterization, for which we obtain a best-fit slope of $0.78\pm0.05$ instead of unity. In contrast, our new MGCR correlator yields a slope of $1.01\pm0.06$, consistent with an intrinsic slope of unity, i.e., a truly linear correlation between $\sigsfr$ and $\gastmff$. 

Our results in Figure~\ref{fig:sflaws} and Table~\ref{tab:fits} strongly suggest that the remaining scatter in the previous $\sigsfr$~vs.~$\gastff$ law by \citet{KDM12} can indeed be primarily attributed to systematic variations in $\mathcal{M}$ as suggested by \citet{Federrath:2013aa}\footnote{See also \citet{KrumholzMcKee2005}, \citet{PadoanNordlund2011}, \citet{HennebelleChabrier2011,HennebelleChabrier2013} and \citet{FederrathKlessen2012} on the detailed derivation of how the SFR depends on the Mach number in the single-freefall versus the multi-freefall formulation.}. The accounting of said Mach number variations in our new SF law has eradicated the discrepancies of up to an order of magnitude, collapsing the scattered relations from \citet{KDM12} and \citet{Federrath:2013aa} into a single, more universal SF law,
\begin{align} \label{eq:newsflaw}
\sigsfr & = 0.4\% \times \gastmff \nonumber \\
          & = 0.4\% \times \gastff \times C_\mathrm{multi-ff} \nonumber \\
           & = 0.4\% \times \gastff \times \left( 1+b^2\mathcal{M}^2\frac{\beta}{\beta+1} \right)^{3/8}.
\end{align}
The reduction in scatter in this SF law strongly indicates that our new model is a physically meaningful descriptor of the SFR.

Thus, our new SF law given by Equation~(\ref{eq:newsflaw}) predicts that the SFR is equal to $\sim\!0.4\%$ of the MGCR in a molecular cloud or galaxy. This value is slightly smaller than the $1\%$ value of the \emph{mean} gas consumption rate, $\gastff$, used in \citet{KDM12}. The reason for this is that the MGCR is naturally larger than the mean gas consumption rate, because we take the density-dependent freefall time into account in our model.

\section{Application of the new model: Mach number predictions for extragalactic sources} \label{sec:appl}

\begin{table}
\caption{Mach number predictions for extragalactic systems obtained by inverting our improved SF law, Equation~(\ref{eq:newsflaw})}
\def\arraystretch{1.6}
\setlength{\tabcolsep}{5.3pt}
\begin{tabular}{lcccc}
\hline
Data Source    & Redshift & Galaxy Type & $\mathcal{M}$ Estimate \\
\hline
\citet{K98}          & $\sim\!0$     & Disk     & $\mathrm{4.0^{ + 2.0}_{ - 1.7}}$\\
                        & $\sim\!0$     & Starburst& $\mathrm{13^{ + 21}_{ - 9.4}}$ \\
\hline
\citet{Daddi:2010aa}       & 1--3 & Disk     & $16^{ + 25}_{ - 11}$ \\
\hline
\citet{Tacconi:2010aa}     & 1--3 & Disk     & $18^{ + 21}_{ - 10}$\\
\hline
\citet{Genzel:2010aa}      & 1--3 & Disk     & $5.7^{ + 0.5}_{ - 0.4}$\\ 
                        & 1--3 & Starburst& $51^{ + 120}_{ - 36} $\\
\hline
\citet{Bouche:2007aa}      & 1--3 & Starburst & $71^{ + 120}_{ - 45}$\\
\hline
\end{tabular}
\begin{minipage}{\linewidth}
\vspace{0.1cm}
\textbf{Notes.} Column~1: Reference. Column~2: Redshift range. Column~3: Galaxy type. Column~4: Geometric mean $\mathcal{M}$ prediction from inverting the improved SF law, Equation~(\ref{eq:newsflaw}), as well as the upper and lower limits of this $\mathcal{M}$ estimate.
\end{minipage}
\label{tab:appl}
\end{table}  

The \citet{KDM12} SF law had utilized extragalactic data of disk and starburst galaxies \citep{K98,Bouche:2007aa,Daddi:2010aa,Genzel:2010aa,Tacconi:2010aa} in addition to Milky Way observations. However, it is difficult to accurately gauge the values of $\mathcal{M}$ for extragalactic sources as we are not able to well resolve individual molecular clouds, especially at high redshifts. $\mathcal{M}$ estimates require measurements of the sound speed and turbulent velocity dispersion of a cloud, which in turn requires high-resolution molecular line data and a temperature measurement. Such data are hard to obtain for molecular clouds in external galaxies, as a complete census of the star-forming molecular cloud population in a given galaxy is needed at sufficiently high resolution to capture the cloud-scale Mach number. There are only very few studies of nearby galaxies (such as M51) which start to resolve molecular clouds in external galaxies \citep{HughesEtAl2013}, but even those do not necessarily yield a complete census of clouds and it is not clear that all cloud properties are converged at the telescope resolutions available to date.

We were hence unable to include extragalactic sources that lack $\mathcal{M}$ estimates in the deduction of our new SF law. They were similarly omitted in our analysis of previous laws in order to arrive at a fair comparison between the different parameterizations of $\sigsfr$ in Figure~\ref{fig:sflaws} and Table~\ref{tab:fits}. Instead, we invert our new model, Equation~(\ref{eq:newsflaw}), in order to compute $\mathcal{M}$ estimates for these extragalactic sources; predictions to be tested by future observations of resolved CO maps in extragalactic sources. Our $\mathcal{M}$ predictions are summarized in Table~\ref{tab:appl}. Since the presence of vigorous star formation in starburst galaxies would result in more violent turbulence of gas compared to disk galaxies, it is expected that the starburst galaxies have greater $\mathcal{M}$ than their disk galaxy counterparts of the same redshift range, which is indeed the systematic trend we find in Table~\ref{tab:appl}. This trend is consistent with the analyses and conclusions presented in \citet{ChabrierEtAl2014}.

\section{Conclusions} \label{sec:conclusions}

We derived a new SF law given by Equation~(\ref{eq:newsflaw}), which is based primarily on the sonic Mach number of turbulence, as a result of the density PDF of molecular, star-forming gas clouds. We find that the SFR is equal to $\sim\!0.4\%$ of the \emph{maximum (multi-freefall) gas consumption rate} (MGCR), which we derived in Section~\ref{sec:theory}. We compared our new model to previous parameterizations of the SFR and determined quantitatively that our new SF law provides a tight linear relation between $\sigsfr$ and MGCR, with a factor of 3--4 less scatter compared to any previous SF law. By inverting Equation~(\ref{eq:newsflaw}), we predict the Mach numbers ($\mathcal{M}$) of the star-forming molecular clouds in extragalactic sources. Our predictions are summarized in Table~\ref{tab:appl}, which anticipate testing via future submillimetre observations.

\acknowledgements
We thank Christopher McKee and the anonymous referee for useful comments, which improved this work. C.F.~acknowledges funding provided by the Australian Research Council's (ARC) Discovery Projects (grants~DP130102078 and~DP150104329). L.J.K.~gratefully acknowledges the support of an ARC~Future~Fellowship and ARC~Discovery~Project~DP130103925.


\begin{thebibliography}{50}
\expandafter\ifx\csname natexlab\endcsname\relax\def\natexlab#1{#1}\fi

\bibitem[{{Berkhuijsen} \& {Fletcher}(2008)}]{BerkhuijsenFletcher2008}
{Berkhuijsen}, E.~M., \& {Fletcher}, A. 2008, \mnras, 390, L19

\bibitem[{{Berkhuijsen} \& {Fletcher}(2015)}]{BerkhuijsenFletcher2015}
---. 2015, \mnras, 448, 2469

\bibitem[{{Bigiel} {et~al.}(2008){Bigiel}, {Leroy}, {Walter}, {Brinks}, {de
  Blok}, {Madore}, \& {Thornley}}]{BigielEtAl2008}
{Bigiel}, F., {Leroy}, A., {Walter}, F., {et~al.} 2008, \aj, 136, 2846

\bibitem[{{Bolatto} {et~al.}(2011){Bolatto}, {Leroy}, {Jameson}, {Ostriker},
  {Gordon}, {Lawton}, {Stanimirovi{\'c}}, {Israel}, {Madden}, {Hony},
  {Sandstrom}, {Bot}, {Rubio}, {Winkler}, {Roman-Duval}, {van Loon},
  {Oliveira}, \& {Indebetouw}}]{BolattoEtAl2011}
{Bolatto}, A.~D., {Leroy}, A.~K., {Jameson}, K., {et~al.} 2011, \apj, 741, 12

\bibitem[{{Bouch{\'e}} {et~al.}(2007){Bouch{\'e}}, {Cresci}, {Davies},
  {Eisenhauer}, {F{\"o}rster Schreiber}, {Genzel}, {Gillessen}, {Lehnert},
  {Lutz}, {Nesvadba}, {Shapiro}, {Sternberg}, {Tacconi}, {Verma}, {Cimatti},
  {Daddi}, {Renzini}, {Erb}, {Shapley}, \& {Steidel}}]{Bouche:2007aa}
{Bouch{\'e}}, N., {Cresci}, G., {Davies}, R., {et~al.} 2007, \apj, 671, 303

\bibitem[{{Brunt}(2010)}]{Brunt2010}
{Brunt}, C.~M. 2010, \aap, 513, A67

\bibitem[{{Chabrier} {et~al.}(2014){Chabrier}, {Hennebelle}, \&
  {Charlot}}]{ChabrierEtAl2014}
{Chabrier}, G., {Hennebelle}, P., \& {Charlot}, S. 2014, \apj, 796, 75

\bibitem[{{Daddi} {et~al.}(2010){Daddi}, {Elbaz}, {Walter}, {Bournaud},
  {Salmi}, {Carilli}, {Dannerbauer}, {Dickinson}, {Monaco}, \&
  {Riechers}}]{Daddi:2010aa}
{Daddi}, E., {Elbaz}, D., {Walter}, F., {et~al.} 2010, \apjl, 714, L118

\bibitem[{{Elmegreen} \& {Scalo}(2004)}]{ElmegreenScalo2004}
{Elmegreen}, B.~G., \& {Scalo}, J. 2004, \araa, 42, 211

\bibitem[{{Evans} {et~al.}(2014){Evans}, {Heiderman}, \&
  {Vutisalchavakul}}]{EvansEtAl2014}
{Evans}, II, N.~J., {Heiderman}, A., \& {Vutisalchavakul}, N. 2014, \apj, 782,
  114

\bibitem[{{Federrath}(2013)}]{Federrath:2013aa}
{Federrath}, C. 2013, \mnras, 436, 3167

\bibitem[{{Federrath} \& {Klessen}(2012)}]{FederrathKlessen2012}
{Federrath}, C., \& {Klessen}, R.~S. 2012, \apj, 761, 156

\bibitem[{{Federrath} {et~al.}(2008){Federrath}, {Klessen}, \&
  {Schmidt}}]{FederrathKlessenSchmidt2008}
{Federrath}, C., {Klessen}, R.~S., \& {Schmidt}, W. 2008, \apjl, 688, L79

\bibitem[{{Federrath} {et~al.}(2010){Federrath}, {Roman-Duval}, {Klessen},
  {Schmidt}, \& {Mac Low}}]{FederrathDuvalKlessenSchmidtMacLow2010}
{Federrath}, C., {Roman-Duval}, J., {Klessen}, R.~S., {Schmidt}, W., \& {Mac
  Low}, M. 2010, \aap, 512, A81

\bibitem[{{Ferri{\`e}re}(2001)}]{Ferriere:2001aa}
{Ferri{\`e}re}, K.~M. 2001, Reviews of Modern Physics, 73, 1031

\bibitem[{{Genzel} {et~al.}(2010){Genzel}, {Tacconi}, {Gracia-Carpio},
  {Sternberg}, {Cooper}, {Shapiro}, {Bolatto}, {Bouch{\'e}}, {Bournaud},
  {Burkert}, {Combes}, {Comerford}, {Cox}, {Davis}, {Schreiber},
  {Garcia-Burillo}, {Lutz}, {Naab}, {Neri}, {Omont}, {Shapley}, \&
  {Weiner}}]{Genzel:2010aa}
{Genzel}, R., {Tacconi}, L.~J., {Gracia-Carpio}, J., {et~al.} 2010, \mnras,
  407, 2091

\bibitem[{{Ginsburg} {et~al.}(2013){Ginsburg}, {Federrath}, \&
  {Darling}}]{GinsburgFederrathDarling2013}
{Ginsburg}, A., {Federrath}, C., \& {Darling}, J. 2013, \apj, 779, 50

\bibitem[{{Gutermuth} {et~al.}(2011){Gutermuth}, {Pipher}, {Megeath}, {Myers},
  {Allen}, \& {Allen}}]{GutermuthEtAl2011}
{Gutermuth}, R.~A., {Pipher}, J.~L., {Megeath}, S.~T., {et~al.} 2011, \apj,
  739, 84

\bibitem[{{Heiderman} {et~al.}(2010){Heiderman}, {Evans}, {Allen}, {Huard}, \&
  {Heyer}}]{HeidermanEtAl2010}
{Heiderman}, A., {Evans}, II, N.~J., {Allen}, L.~E., {Huard}, T., \& {Heyer},
  M. 2010, \apj, 723, 1019

\bibitem[{{Hennebelle} \& {Chabrier}(2011)}]{HennebelleChabrier2011}
{Hennebelle}, P., \& {Chabrier}, G. 2011, \apjl, 743, L29

\bibitem[{{Hennebelle} \& {Chabrier}(2013)}]{HennebelleChabrier2013}
---. 2013, \apj, 770, 150

\bibitem[{{Heyer} \& {Brunt}(2012)}]{HeyerBrunt2012}
{Heyer}, M.~H., \& {Brunt}, C.~M. 2012, \mnras, 420, 1562

\bibitem[{{Hopkins}(2013)}]{Hopkins2013IMF}
{Hopkins}, P.~F. 2013, \mnras, 430, 1653

\bibitem[{{Hughes} {et~al.}(2013){Hughes}, {Meidt}, {Schinnerer}, {Colombo},
  {Pety}, {Leroy}, {Dobbs}, {Garc{\'{\i}}a-Burillo}, {Thompson}, {Dumas},
  {Schuster}, \& {Kramer}}]{HughesEtAl2013}
{Hughes}, A., {Meidt}, S.~E., {Schinnerer}, E., {et~al.} 2013, \apj, 779, 44

\bibitem[{{Kainulainen} {et~al.}(2009){Kainulainen}, {Beuther}, {Henning}, \&
  {Plume}}]{KainulainenEtAl2009}
{Kainulainen}, J., {Beuther}, H., {Henning}, T., \& {Plume}, R. 2009, \aap,
  508, L35

\bibitem[{{Kainulainen} {et~al.}(2013){Kainulainen}, {Federrath}, \&
  {Henning}}]{KainulainenFederrathHenning2013}
{Kainulainen}, J., {Federrath}, C., \& {Henning}, T. 2013, \aap, 553, L8

\bibitem[{{Kainulainen} {et~al.}(2014){Kainulainen}, {Federrath}, \&
  {Henning}}]{Kainulainen:2014aa}
---. 2014, Science, 344, 183

\bibitem[{{Kennicutt} \& {Evans}(2012)}]{KennicuttEvans2012}
{Kennicutt}, R.~C., \& {Evans}, N.~J. 2012, \araa, 50, 531

\bibitem[{{Kennicutt}(1998)}]{K98}
{Kennicutt}, Jr., R.~C. 1998, \apj, 498, 541

\bibitem[{{Krumholz}(2014)}]{Krumholz2014}
{Krumholz}, M.~R. 2014, Physics Reports, in press (arXiv:1402.0867)

\bibitem[{{Krumholz} {et~al.}(2012){Krumholz}, {Dekel}, \& {McKee}}]{KDM12}
{Krumholz}, M.~R., {Dekel}, A., \& {McKee}, C.~F. 2012, \apj, 745, 69

\bibitem[{{Krumholz} \& {McKee}(2005)}]{KrumholzMcKee2005}
{Krumholz}, M.~R., \& {McKee}, C.~F. 2005, \apj, 630, 250

\bibitem[{{Lada} {et~al.}(2010){Lada}, {Lombardi}, \&
  {Alves}}]{LadaLombardiAlves2010}
{Lada}, C.~J., {Lombardi}, M., \& {Alves}, J.~F. 2010, \apj, 724, 687

\bibitem[{{Leroy} {et~al.}(2008){Leroy}, {Walter}, {Brinks}, {Bigiel}, {de
  Blok}, {Madore}, \& {Thornley}}]{LeroyEtAl2008}
{Leroy}, A.~K., {Walter}, F., {Brinks}, E., {et~al.} 2008, \aj, 136, 2782

\bibitem[{{Mac Low} \& {Klessen}(2004)}]{MacLowKlessen2004}
{Mac Low}, M.-M., \& {Klessen}, R.~S. 2004, RvMP, 76, 125

\bibitem[{{McKee} \& {Ostriker}(2007)}]{McKeeOstriker2007}
{McKee}, C.~F., \& {Ostriker}, E.~C. 2007, \araa, 45, 565

\bibitem[{{Molina} {et~al.}(2012){Molina}, {Glover}, {Federrath}, \&
  {Klessen}}]{MolinaEtAl2012}
{Molina}, F.~Z., {Glover}, S.~C.~O., {Federrath}, C., \& {Klessen}, R.~S. 2012,
  \mnras, 423, 2680

\bibitem[{{Padoan} {et~al.}(2014){Padoan}, {Federrath}, {Chabrier}, {Evans},
  {Johnstone}, {J{\o}rgensen}, {McKee}, \& {Nordlund}}]{PadoanEtAl2014}
{Padoan}, P., {Federrath}, C., {Chabrier}, G., {et~al.} 2014, Protostars and
  Planets VI, 77

\bibitem[{{Padoan} \& {Nordlund}(2011)}]{PadoanNordlund2011}
{Padoan}, P., \& {Nordlund}, {\AA}. 2011, \apj, 730, 40

\bibitem[{{Price} {et~al.}(2011){Price}, {Federrath}, \&
  {Brunt}}]{PriceFederrathBrunt2011}
{Price}, D.~J., {Federrath}, C., \& {Brunt}, C.~M. 2011, \apjl, 727, L21

\bibitem[{{Rathborne} {et~al.}(2014){Rathborne}, {Longmore}, {Jackson},
  {Kruijssen}, {Alves}, {Bally}, {Bastian}, {Contreras}, {Foster}, {Garay},
  {Testi}, \& {Walsh}}]{RathborneEtAl2014}
{Rathborne}, J.~M., {Longmore}, S.~N., {Jackson}, J.~M., {et~al.} 2014, \apjl,
  795, L25

\bibitem[{{Renaud} {et~al.}(2012){Renaud}, {Kraljic}, \&
  {Bournaud}}]{RenaudEtAl2012}
{Renaud}, F., {Kraljic}, K., \& {Bournaud}, F. 2012, \apjl, 760, L16

\bibitem[{{Schmidt}(1959)}]{Schmidt1959}
{Schmidt}, M. 1959, \apj, 129, 243

\bibitem[{{Schneider} {et~al.}(2012){Schneider}, {Csengeri}, {Hennemann},
  {Motte}, {Didelon}, {Federrath}, {Bontemps}, {Di Francesco}, {Arzoumanian},
  {Minier}, {Andr{\'e}}, {Hill}, {Zavagno}, {Nguyen-Luong}, {Attard},
  {Bernard}, {Elia}, {Fallscheer}, {Griffin}, {Kirk}, {Klessen}, {K{\"o}nyves},
  {Martin}, {Men'shchikov}, {Palmeirim}, {Peretto}, {Pestalozzi}, {Russeil},
  {Sadavoy}, {Sousbie}, {Testi}, {Tremblin}, {Ward-Thompson}, \&
  {White}}]{SchneiderEtAl2012}
{Schneider}, N., {Csengeri}, T., {Hennemann}, M., {et~al.} 2012, \aap, 540, L11

\bibitem[{{Schneider} {et~al.}(2013){Schneider}, {Andr{\'e}}, {K{\"o}nyves},
  {Bontemps}, {Motte}, {Federrath}, {Ward-Thompson}, {Arzoumanian},
  {Benedettini}, {Bressert}, {Didelon}, {Di Francesco}, {Griffin}, {Hennemann},
  {Hill}, {Palmeirim}, {Pezzuto}, {Peretto}, {Roy}, {Rygl}, {Spinoglio}, \&
  {White}}]{SchneiderEtAl2013}
{Schneider}, N., {Andr{\'e}}, P., {K{\"o}nyves}, V., {et~al.} 2013, \apjl, 766,
  L17

\bibitem[{{Schruba} {et~al.}(2011){Schruba}, {Leroy}, {Walter}, {Bigiel},
  {Brinks}, {de Blok}, {Dumas}, {Kramer}, {Rosolowsky}, {Sandstrom},
  {Schuster}, {Usero}, {Weiss}, \& {Wiesemeyer}}]{SchrubaEtAl2011}
{Schruba}, A., {Leroy}, A.~K., {Walter}, F., {et~al.} 2011, \aj, 142, 37

\bibitem[{{Tacconi} {et~al.}(2010){Tacconi}, {Genzel}, {Neri}, {Cox}, {Cooper},
  {Shapiro}, {Bolatto}, {Bouch{\'e}}, {Bournaud}, {Burkert}, {Combes},
  {Comerford}, {Davis}, {Schreiber}, {Garcia-Burillo}, {Gracia-Carpio}, {Lutz},
  {Naab}, {Omont}, {Shapley}, {Sternberg}, \& {Weiner}}]{Tacconi:2010aa}
{Tacconi}, L.~J., {Genzel}, R., {Neri}, R., {et~al.} 2010, \nat, 463, 781

\bibitem[{{V{\'a}zquez-Semadeni}(1994)}]{Vazquez1994}
{V{\'a}zquez-Semadeni}, E. 1994, \apj, 423, 681

\bibitem[{{Wu} {et~al.}(2010){Wu}, {Evans}, {Shirley}, \& {Knez}}]{WuEtAl2010}
{Wu}, J., {Evans}, II, N.~J., {Shirley}, Y.~L., \& {Knez}, C. 2010, \apjs, 188,
  313

\bibitem[{{Yusef-Zadeh} {et~al.}(2009){Yusef-Zadeh}, {Hewitt}, {Arendt},
  {Whitney}, {Rieke}, {Wardle}, {Hinz}, {Stolovy}, {Lang}, {Burton}, \&
  {Ramirez}}]{YusefZadeh2009}
{Yusef-Zadeh}, F., {Hewitt}, J.~W., {Arendt}, R.~G., {et~al.} 2009, \apj, 702,
  178

\end{thebibliography}

\end{document}